\documentclass[12pt,preprint]{aastex}

\newcommand{\be}{\begin{eqnarray}}
\newcommand{\ee}{\end{eqnarray}}

\newcommand{\smass}{{M_{\odot}}}

\shorttitle{Triple Blackholes II}
\shortauthors{Iwasawa et al. }

\begin{document}

\title{Evolution of Massive Blackhole Triples II --- 
The effect of the BH triples dynamics 
on the structure of the galactic nuclear}

\author{Masaki Iwasawa \altaffilmark{1,2} , Yoko Funato\altaffilmark{1}
and Junichiro Makino\altaffilmark{2}
}
\email{iwasawa@cfca.jp}

\altaffiltext{1}{Department of General System Studies, University of Tokyo,
3-8-1 at Komaba, Komaba, Meguro-ku, Tokyo 153-8902, Japan}
\altaffiltext{2}{Division of Theoretical Astronomy, National
Astronomical Observatory, 2-2-1 Osawa, Mitaka, Tokyo 181-8588, Japan}

\begin{abstract}

In this paper, we investigate the structures of galaxies
which either have or have had three BHs using $N$-body simulations,
and compare them with those of galaxies with binary BHs.
We found that the cusp region of a galaxy which have (or had) triple
BHs is significantly larger and less dense than that of a
galaxy with binary BHs of the same mass. Moreover, the size of the
cusp region depends strongly on the evolution history of triple BHs,
while in the case of binary BHs, the size of the cusp is
determined by the mass of the BHs.  In  galaxies which have (or had)
three BHs, there is a region with significant radial velocity
anisotropy, while such a region is not observed in galaxies with
binary BH. These differences come from the fact that with triple BHs
the energy deposit to the central region of the galaxy can be much
larger due to multiple binary-single BH scatterings.
Our result suggests that we can discriminate between
galaxies which experienced triple BH interactions with those which did
not, through the observable signatures such as the cusp size and
velocity anisotropy.

\end{abstract}


\section{Introduction}

According to recent observations, elliptical galaxies can be
classified into two groups: ``weak-cusp'' galaxies and ``strong-cusp''
galaxies \citep{lauer95, faber97}.  The central surface brightness
profiles of the weak-cusp galaxies are expressed as
 $\Sigma(R) \propto R^{-\gamma}$ 
with $\gamma \le 0.3$, and those of the strong-cusp
galaxy the same formula with $\gamma \ge 0.5$. The slope of the volume
density profile of strong-cusp galaxies is around $-2$, being
consistent with the isothermal cusp. Such a steep cusp is naturally
formed in dissipative process involving gas dynamics and star
formation. On the other hand, the weak cusp corresponds to the slope
of the volume density shallower than $-1$, 
which is not likely to be formed through dissipative process.

One possible way to form a weak cusp is the merging of two galaxies
containing black holes (BHs). When two galaxies, each with a
black hole, merge, two BHs sink toward the center of the merger
remnant  to form a binary through the dynamical friction
\citep{bbr80}.  The back reaction of the dynamical friction heats up
field stars. As a result, a shallow cusp of stars develops in
the central region \citep{Ebisuzakietal1991,NM99,merritt06}.

One problem with this binary BH
scenario is what would be the final fate of the binary BH.
\citet{bbr80} pointed out that the merging timescale of the binary BH
might be very long, after the binary BH ejected out the stars which
can interact with the binary (loss cone depletion).
The stars will be supplied in the timescale of the relaxation time,
which is much longer than the Hubble time.
Recent $N$-body simulations confirmed this
theoretical estimate \citep{MF04, berczik2005}.

If galaxies are formed through hierarchical clusterings, in many
cases, a binary BH is formed after a merger event. If there were
sufficient gas left in the merger, interaction with gas might lead
to the quick merging of the two BHs. However, in the case of ``dry''
mergers which would result in the formation of giant ellipticals, by
definition not much gas is left and it would be difficult for two BHs
to merge.

If one galaxy with binary BH and the other with single BH merge, the
central BHs form a triple system.  Iwasawa, Funato \& Makino (2006,
hereafter referred to as Paper I) investigated the evolution of triple
BH system in the galactic center, using $N$-body simulations.  They
found that the strong binary-single BH interaction \citep{ME94} and
the Kozai cycle \citep{Kozai62, BLS} drives the eccentricity of
the BH binary high enough that two BHs merge
quickly through gravitational wave radiation.
\cite{HL07} performed statistical simulations of evolution of
central BH systems and reached a similar conclusion.

This paper is a follow-up of Paper I.
In this paper, we investigate
the structure of a galaxy containing three BHs.
We also investigated their observational properties
which would help us to find galaxies which have (or had) triple BHs.

We performed $N$-body simulations of the evolution of triple (or binary)
BH in a host galaxy, in order to study the dynamical
evolution of the structure of stellar systems containing BHs.  In our
simulations, both dynamical evolution of BHs and that of field stars are
integrated consistently.  The interaction between BHs affects not only
the evolution of themselves but also the spatial and kinematic
structure of field stars around them.  In turn, the distribution of
field stars  affects the interaction between BHs. 
To understand the structure of galaxies containing BHs, a  self-consistent
simulation in which the orbits of BHs and stars are treated
self-consistently is essential. 

The structure of this paper is as follows.  In section 2, we describe
the initial models and the method of our numerical simulations.  In
section 3, we show the effect of the BH triples dynamics
on the structure of the galaxy.
Summary and discussion are given in section 4.

\section{Initial Models and Numerical Methods}

\subsection{Initial Models}

The initial setup of the models is basically the same as that in Paper
I.  
For the galaxy model, we used a King model with $W_{0}=11$, where
$W_{0}$ is the nondimensional central potential of King models
\citep{kin66, GD}.  We adopted the standard $N$-body units
\citep{Hg}, 
in which $m_{{\mathrm gal}}=1.0, E_{{\mathrm gal}}=-0.25,G=1$.
Here, $m_{{\mathrm gal}}$ and $E_{{\mathrm gal}}$ are the total
mass and total binding energy of the galaxy not including BHs
and $G$ is the gravitational constant. 
We placed three equal-mass BH particles in $N$-body models of a
spherical galaxy.  Two of three BHs particles are initially placed
at the position $(\pm 0.005, 0.0, 0.0)$ with velocity $(0.0, \pm 0.15,
0.0)$, and the third one is at the position $(0.1, 0.0, 0.0)$ with
velocity $(0.0,0.0,0.0)$.  In order to compare the structure of a
galaxy with three BHs with that of a galaxy with two BHs, we also
performed simulations of galaxies with two BHs.  Total mass of BHs in
two-BH models was kept to be the same as that in the corresponding
three-BH models.  For two-BH model, two BHs are placed at the position
$(\pm 0.1, 0.0, 0.0)$ with velocity $(0.0, \pm 0.1, 0.0)$.  The
quantitative properties of models and initial conditions of our
simulation is summarized in table \ref{tbl1}.

We set the number of FS particles $N_{\mathrm FS}=262144$
in all simulations.  
In order to investigate the effect of triple BHs on the galaxy, we
set the mass ratio between galaxy and total mass of BHs to,
${m_{\mathrm{BH,tot}}/m_{\mathrm{gal}}} \sim 0.003$, 
where $m_{\mathrm{BH,tot}}$ is total
mass of BH, in our standard model.
This value is what is suggested by 
recent observations \citep{KR95, magorrian98, marconi03}.
Table \ref{tbl2} shows all models.

\begin{deluxetable}{lcc}
\tablecaption{Model Parameters \label{tbl1}}
\tablewidth{0pt}
\tablehead{\colhead{Parameter} & \colhead{Symbol}   & \colhead{Value}}
\startdata
Mass of galaxy &$m_{\mathrm{gal}}$ &$1.0$ \\
Mass of FS &$m_{\mathrm{FS}}$ &$1/262144$  \\
Mass of BH &$m_{\mathrm{BH}}$ &$0.001 - 0.0045$  \\
Total mass of BHs &$m_{\mathrm{BH,tot}}$ &$0.003 - 0.009$  \\
Number of FS &$N_{\mathrm{FS}}$ &$262144$ \\
Number of BH &$N_{\mathrm{BH}}$ &$2 \rm{or} 3$ \\
Gravitational constant &$G$ &$1$ \\ 
Total energy &$E_{\mathrm{gal}}$ &$-0.25$ \\
Softening between BHs &$\epsilon_{\mathrm{BH-BH}}$ &$0.0$ \\
Softening between field stars and BH &$\epsilon_{\mathrm{FS-BH}}$ &$10^{-7}$ \\
Softening between field stars &$\epsilon_{\mathrm{FS-FS}}$ &$0.001$ \\
\enddata
\end{deluxetable}

\begin{deluxetable}{ccccccc}
\tablecaption{Model List \label{tbl2}}
\tablewidth{0pt}
\tablehead{
\colhead{Model name} & \colhead{$N_{\mathrm{FS}}$} & \colhead{$N_{\mathrm{BH}}$}
 & \colhead{$W_0$}  & \colhead{$m_{\mathrm{BH}}$}  & \colhead{final state}
}
\startdata
    M1T &262144 &3 &11 &0.001 &merge \\
    M2T &262144 &3 &11 &0.002 &merge \\
    M3T &262144 &3 &11 &0.003 &merge \\ \hline
    M1B &262144 &2 &11 &0.0015 &no merge \\
    M2B &262144 &2 &11 &0.003 &no merge \\
    M3B &262144 &2 &11 &0.0045 &no merge \\
\enddata
\end{deluxetable}

\subsection{Numerical Method}

The numerical method is the same as that we used in  Paper I.
To summarize, we used a softened Newton gravity for the forces between
field stars and those between each BH particle and field stars,
while for forces between BH particles 
we applied post-Newtonian approximation  to
include the back reaction of
gravitational radiation to the BHs.
For the term corresponding to the radiation of the gravitational wave,
we used approximately 2.5-order post-Newtonian approximation
\citep{Damour87}.

We used the Plummer softening and set the value
$\epsilon_{{\rm FS-FS}} = 10^{-4}$, $\epsilon_{{\rm FS-BH}} = 10^{-7}$
and $\epsilon_{{\rm BH-BH}} = 0.0$,
where $\epsilon_{{\rm XX-YY}}$ is softening parameters
used for a XX particle - a YY particle interaction.
Time integration scheme is the 4-th order Hermite
scheme \citep{her} with individual variable time steps.

In order to calculate the acceleration due to field particles, we used
GRAPE-6 \citep{grape6}, the special-purpose computer for the
gravitational $N$-body problem.  Forces from BH particles, both
Newtonian and gravitational wave terms, are calculated on the host
computer.  In all runs, the energy (corrected for the loss of energy
through GW radiation) is conserved  better than  0.2 $\%$

\subsection{Physical Scales}

The mass unit in our simulation corresponds to $10^{11} \smass$.
Thus, the mass of each BH particle is $10^{8} \smass$ in model M1T.
We assume that the velocity dispersion of the initial King model at the
King radius corresponds to 300 km/s.  In other words, we set the light
velocity $c$ in the $N$-body unit to 1006.  
The unit of length and time are about 4.9kpc and
16Myr, respectively.  The king radius is about $100 \rm{pc}$.

\section{Result}

\subsection{Evolution of BH System}

\begin{figure}
\plotone{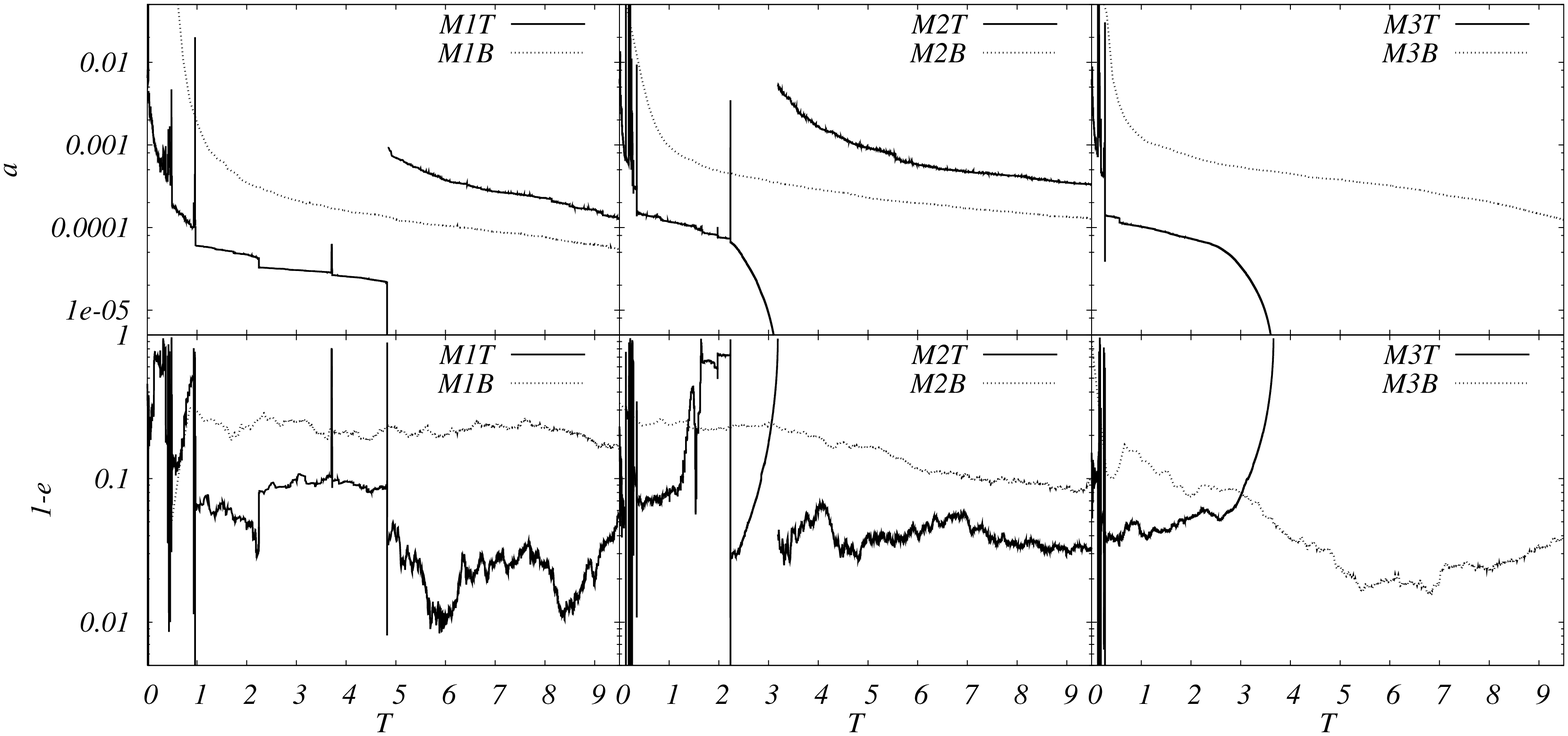}
\caption{
Evolution of semi-major axis (top) and eccentricity (bottom) of the most
 strongly bound pair of BHs for all models.
Solid and doted curves indicate the results of models  MxT (triple BH)
and MxB (binary BH) respectively.
\label{fig1}
}
\end{figure}

Figure \ref{fig1}  shows the
evolution of semi-major axis $a$ and eccentricity $e$ of BH binaries in
all models.   For  models with
triple BHs (model name MxT), the definition of the binary of BHs is
the most strongly bound pair of three BHs.  For the model M1T, jumps
in $a$ and $e$ at $T \sim 0.45, 0.96, 2.2, 3.7$ and $4.8$ are results
of interactions between the BH binary and the third BH.
At $T\sim4.8$, two BHs merged.  
This merging is driven by the impulsive eccentricity
growth caused by a binary-single interaction (Paper I), followed by
the spiral-in due to gravitational wave radiation.  The merging
criterion we used here is that the distance between two BHs becomes
smaller than three times of the Schwarzschild radius of a BH.  We
replaced the merged BHs by a new single BH.  This new BH has the same
mass, position and velocity as those of the barycenter of original
binary.

As shown in Figure \ref{fig1}, the merged BH and the third one form a
binary in model M1T.  In models  M1B and M1T, after $T=4.8$, $a$ and $e$
change slowly, and any BH merger  does not occur until
$T=10$. In both models, a binary BH is left around the center of the
host galaxy. Note that this slow shrink of the BH binary orbit is due
to the refilling of the loss cone through two-body relaxation, which
is enhanced because of the relatively small number of particles we used
($N_{{\rm FS}}=262144$). In real galaxies the timescale of orbital evolution of binary
BH would be much longer after the loss cone is depleted. 

In model M2T, the overall behavior of BHs is rather similar to
that in model M1T. After strong triple interactions at $T=0.35$ 
and $2.2$, the binary merged through gravitational wave radiation. On
the other hand, in model M3T, after the first triple interaction at
$T=0.3$, the binary BH hardens through the interaction with field
stars and finally merged through gravitational wave radiation. The
difference between these three models is at least partly just a
coincidence, since the final result depends on the outcome of the
first triple encounter at time around 0.4. 

On the other hand, the behavior of models with binary BHs are all very
similar, showing slow hardening driven by the loss-cone refill through
two-body relaxation.

\subsection{Density Structure}

\subsubsection{Density Profile}

\begin{figure}
\plottwo{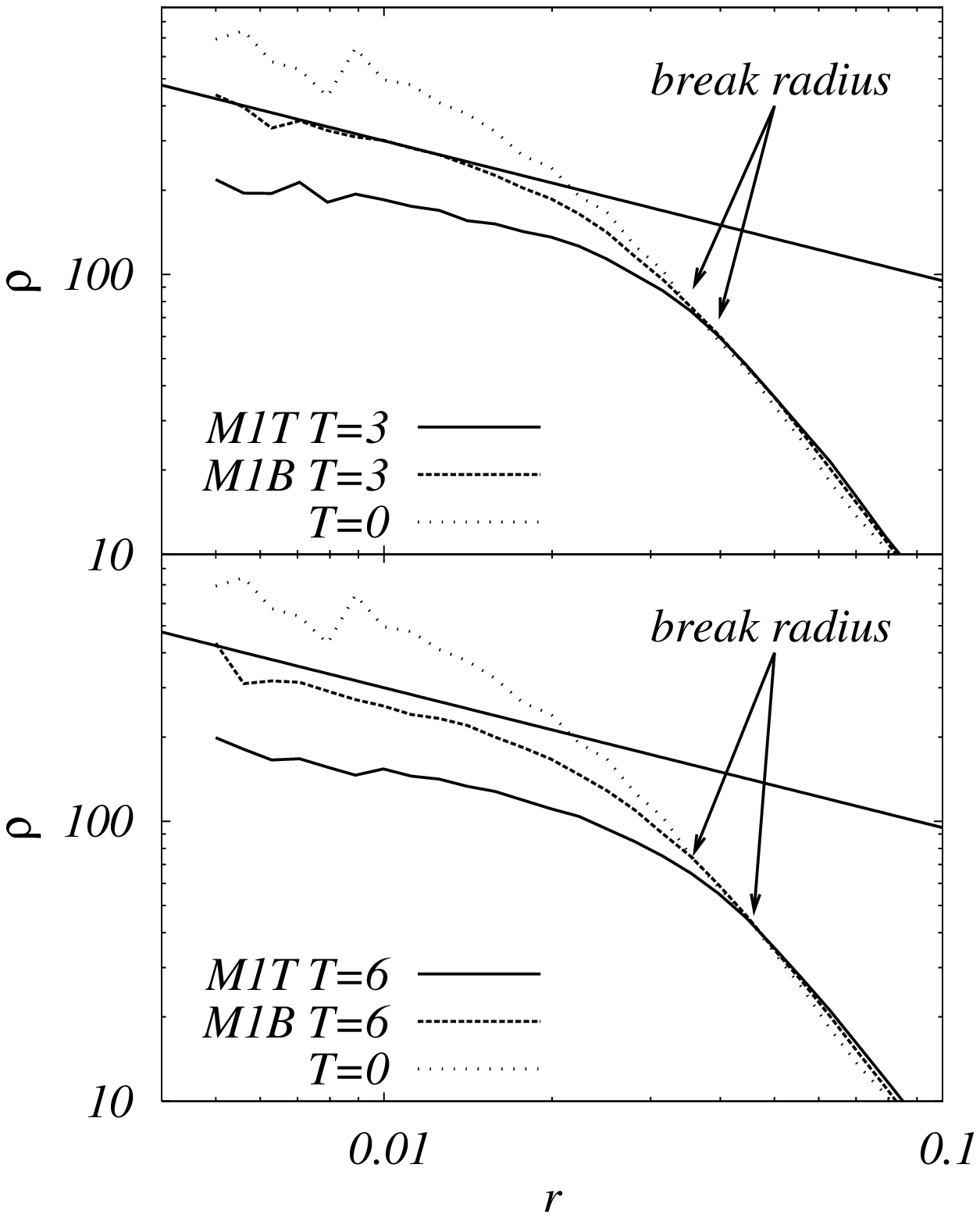}{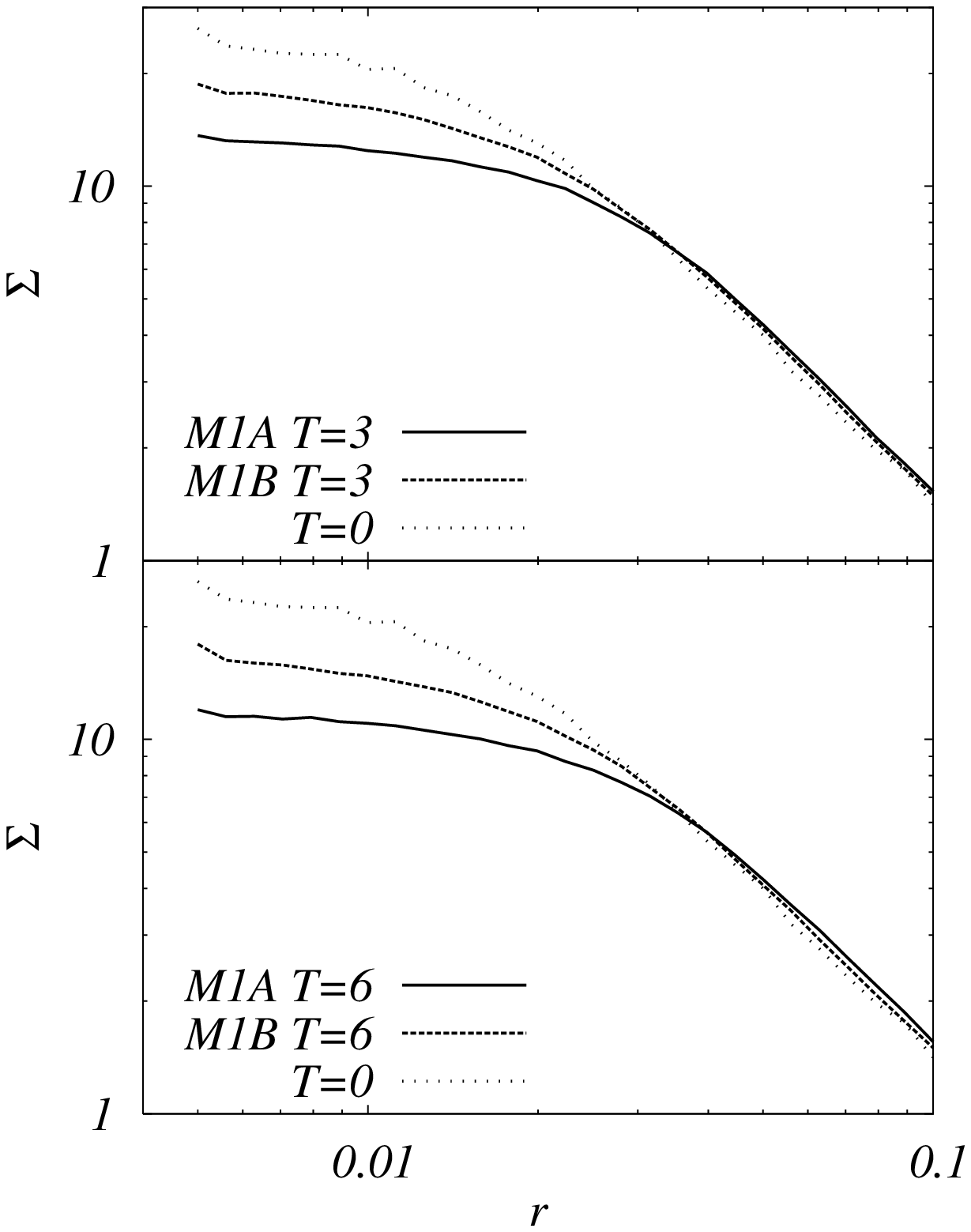}
\caption{
Volume (left) and surface (right) density profiles 
for models M1T (solid) and M1B (dashed).
Dotted curves show the initial profile.
Solid lines indicate the power low $\rho \propto r^{-0.5}$.
Top panel shows the profiles at $T=3$,
and bottom shows those at $T=6$.
\label{fig2}
}
\end{figure}

In Figure \ref{fig2}, we show the spatial (left) and surface (right)
density profiles of models M3T and M3B at $T=3$ and $T=6$.  In both
models the central density decreases and a weak cusp develops in the
central region by $T=3$.  They do not change significantly after
$T=3$.  This cusp is expressed as $\rho \propto r^{-\gamma}$ , where
$\gamma \sim 0.5$.  For both models the slope of cusps can be
explained by the simple theory of \citet{NM99}.

From figure \ref{fig2} we can see that there is a clear difference
between models M1T and M1B. The size of the cusp region is bigger for
M1T and the density is lower.  This result indicates that a triple BH
system is more efficient in heating the central region and ejecting
the stars than a BH binary is.

In the case of a galaxy with a binary BH, a BH binary hardens in a
monotonic fashion through interactions with FSs. In a galaxy with a
triple BH system, a BH binary experiences multiple interactions with a
single BH. In each event a single BH and the center of mass motion of
binary BH acquires energy.  This energy is then transferred to field
stars through dynamical friction.  As a result, the field stars are heated up
and the cusp becomes wider. This mechanism is shown in Figure
\ref{fig3} more clearly.

\subsubsection{Evolution of Cusp Region}

\begin{figure}
\epsscale{1.0}
\plotone{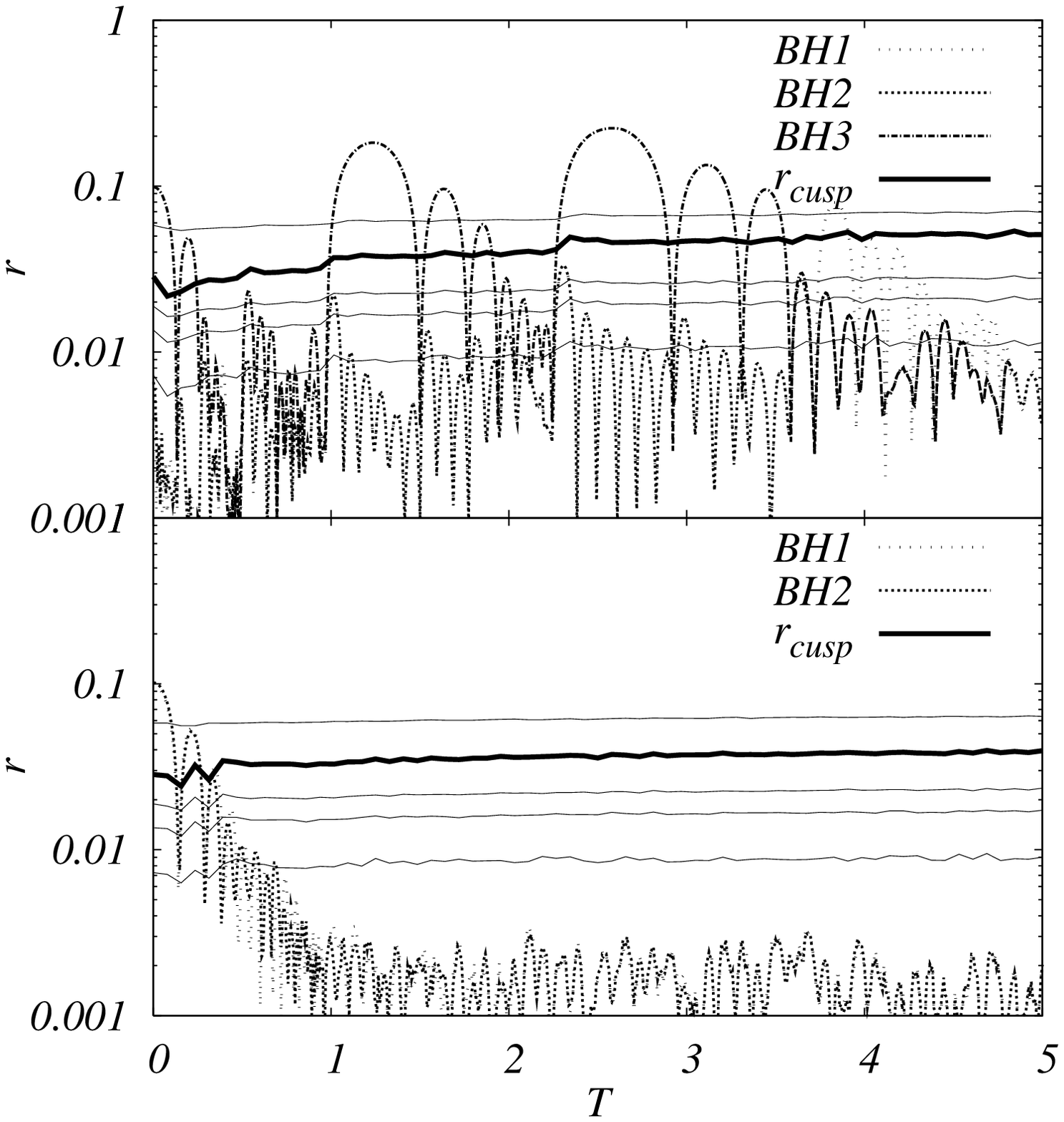}
\caption{
 Evolution of Lagrangian radii, 
 radial distance of BHs
 from the center of the galaxy,
 and the cusp radius,
 for models M1T(top panel) and M1B(bottom panel).
 The four solid curves show Lagrange radii
 ($0.1\%$, $0.5\%$, $1\%$, $5\%$ of total mass).
 Thick dashed, dotted and dash-dotted curves show 
 the radial distances from the galactic center for BHs.
 Thick solid curves show the cusp radius.
\label{fig3}
}

\end{figure}

\begin{figure}
\plottwo{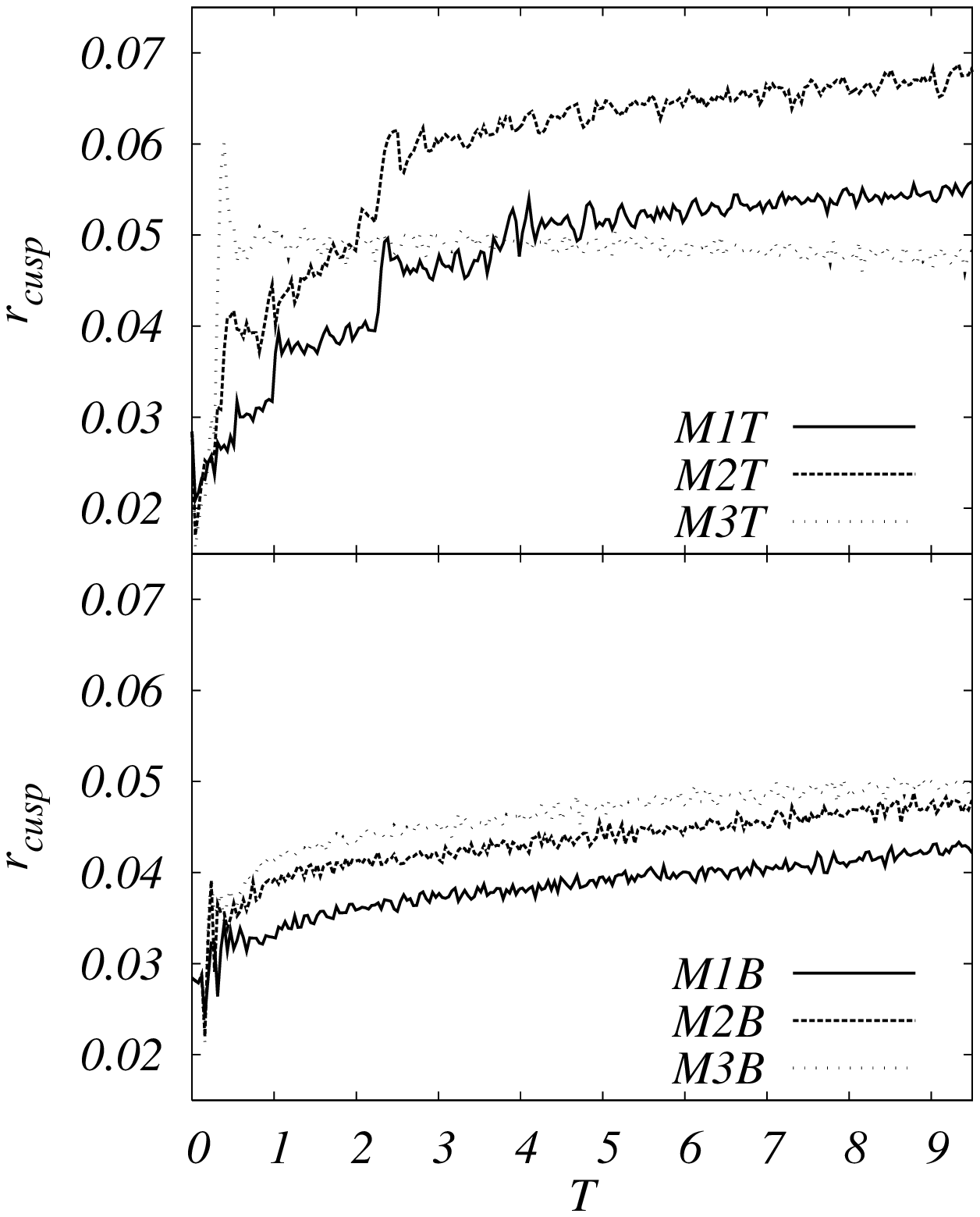}{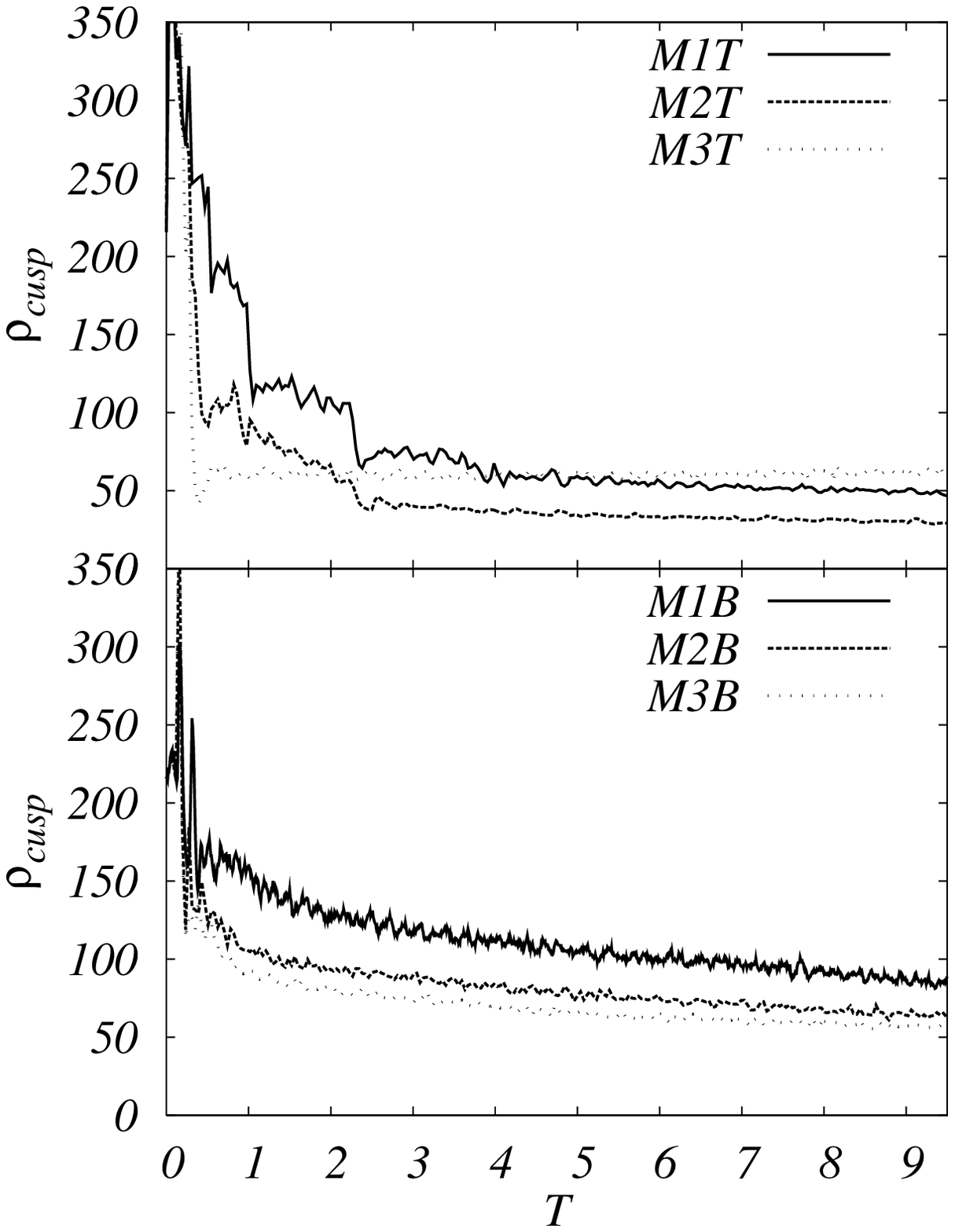}
\caption{
Evolution of cusp radii (left) and cusp densities (right)
for all models.
Top and bottom panels show the results for triple and two BH models respectively.
Solid, dashed, and dotted curves represent 
the results of models M1x, M2x and M3x.
\label{fig4}
}
\end{figure}

In figures \ref{fig3} and \ref{fig4}, time evolution of the Lagrangian
radii of field stars, the cusp radius and radial positions of BHs for all
models are plotted.

Following  \citet{CH85}, we defined the cusp radius as
\begin{eqnarray}
 r_{cusp} &\equiv& \frac{\sum_i \rho_i r_i}{\sum_i \rho_i}, 
\end{eqnarray}
where $\rho_i$ is the local mass density around the field particle
with index $i$ and
the summation is done over field particles. The local mass density is
defined as
\begin{eqnarray}
 \rho_i &=&\frac{m_{6,i}}{\frac{4}{3} \pi r^3_{6,i}},
\end{eqnarray}
where, $m_{6,i}$ is the total FS particle mass contained within
$r_{6,i}$ which is the distance from the field particle $i$ to its sixth
nearest neighbor. 

In top panel of figure \ref{fig3}, the evolution of model M1T is
shown. We can see that the Lagrangian radii and cusp radius change
rather impulsively at the same time as the strong triple scattering
events occur and BHs are ejected out of the central region. This
expansion is because of the indirect heating due to the removal of
massive BH particles from the center. After the scattering events, BHs
sink towards the center of the system through dynamical friction, and
the Lagrangian radii show small expansion. Thus, the expansion is
driven by the BH particles. 
In the case of model M1B, the expansion is smooth and smaller than
that in model M1T.

Figure \ref{fig4} shows the evolution of the cusp radius 
and the mean cusp density of host galaxies for all models.  We can
clearly see that the expansion of the cusp radius in models MxT are
driven by triple scattering events. In the case of model M3T, there
was only one scattering event and therefore only one rapid expansion
event for the cusp radius. In this model, the third BH is ejected out
of the galaxy and never returned.
Thus, one merged BH
particle is left at the center of the galaxy. Slow decrease of the
cusp radius is due to the thermal evolution of the field star system.

\begin{figure}
\plotone{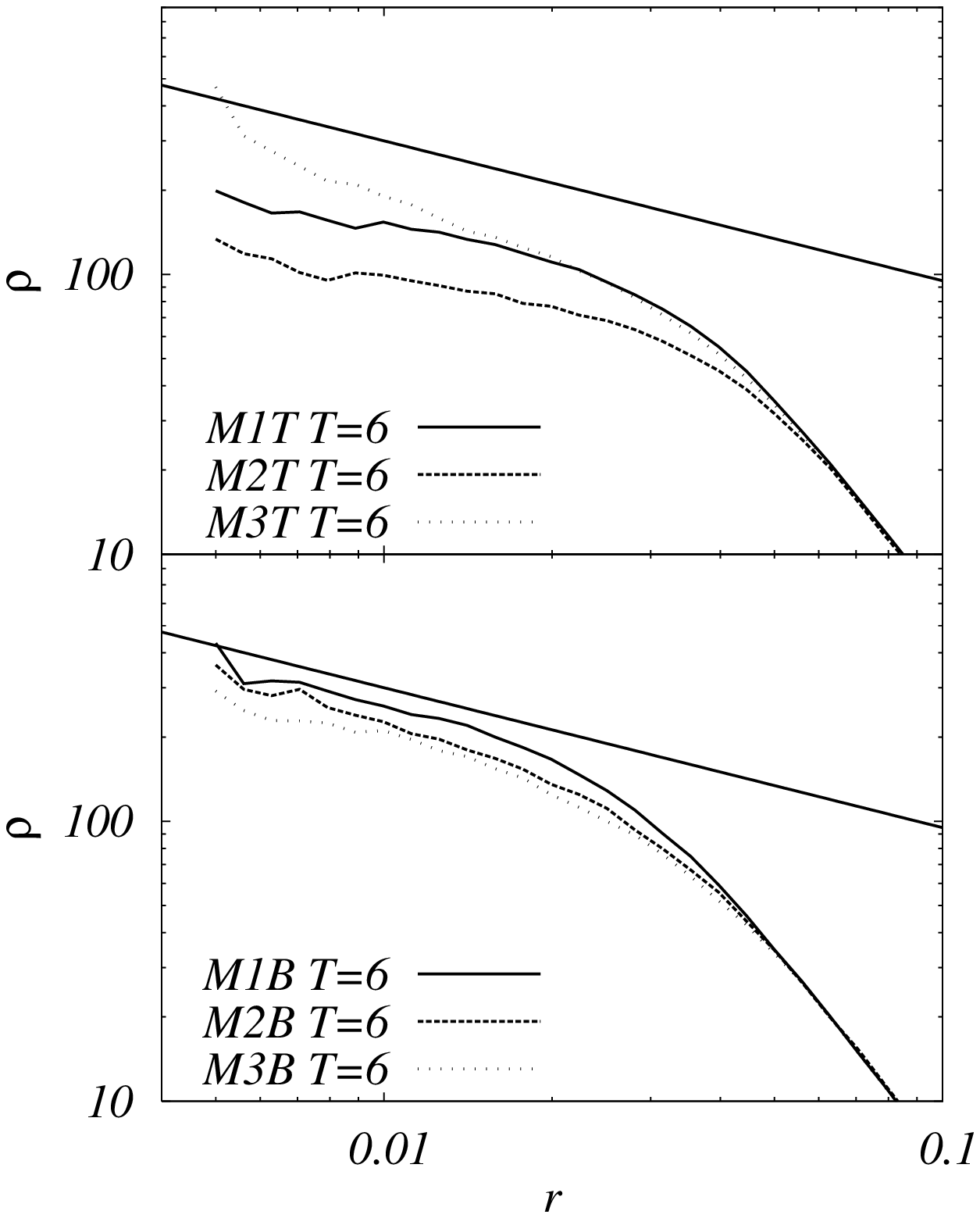}
\caption{
Spatial density profiles for all runs at $T=6$.
Solid lines indicate the power law $\rho \propto r^{-0.5}$.
Top panel shows the results of three-BH models,
and bottom shows those of two-BH models.
\label{fig5}
}
\end{figure}

Figure \ref{fig5} shows the density profile for all runs at $T=6$.
The density profile for all models are similar in their shapes.
The density profile of model M3T is a bit steeper than those of other
models.
This is because only one binary-single scattering event 
is not enough to make the density profile
$\rho \propto r^{-0.5}$ .

For models with triple BHs (MxT),
the radius and density of the cusp
are determined not only by 
the total mass of BHs, but also by the number of binary-single 
BH scattering events.
On the other hand, in the case of binary BHs, the total  mass of BHs
determines the structure of the cusp. 
In addition, in all cases the
cusps in models MxT are bigger than those in models MxB of the
corresponding BH mass.

\subsection{Mass deficit}

The discussion in the previous subsection can be re-interpreted in
terms of the so-called ``mass deficit''
(Milosavljevi{\' c} \& Merritt 2001, hereafter referred to as MM2001 ).
They argued that one can estimate the mass removed by BH
binary activity by comparing the singular isothermal profile and an actual
cusp profile, and they argued that the difference, which they named
``mass deficit'', retains the memory of the merging history of the
black hole. \citet{Graham04} argued that MM2001's method to estimate the
mass deficit would give too large values like $m_{{\rm def}} \sim 10m_{{\rm BH}}$,
where $m_{{\rm def}}$ is the mass deficit, while a more reasonable method
would give $m_{{\rm def}} \sim m_{{\rm BH}}$. \citet{merritt06} claimed that in repeated
mergers $m_{{\rm def}} \sim Nm_{{\rm BH}}$, where $N$ is the number of merging
events, though his simulation result seems to suggest that such linear
scaling does not hold for $N>3$. In addition, simulations of the repeated
merger of \citet{merritt06} ignored the fact that additional BHs in
reality comes with their host galaxies. 
\citet{ME96}
have already shown that in the case of hierarchical repeated mergings,
the density profile of merger remnant converge to a universal
profile. In other words, $m_{{\rm def}}/m_{{\rm BH}}$ does not tell much about
the past merger history.

\begin{figure}
\plotone{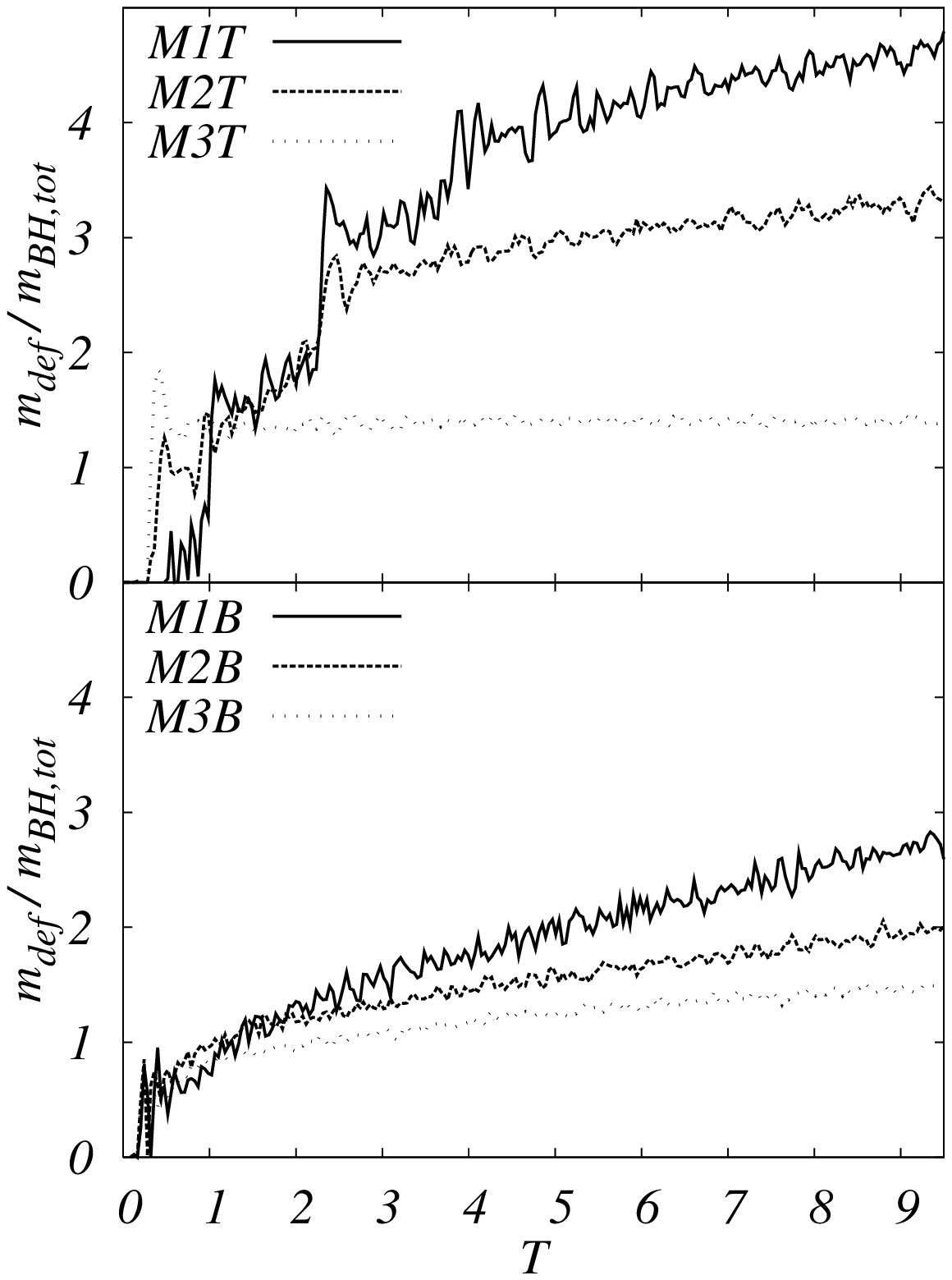}
 \caption
 {
Time evolution of $m_{def}$ normalized by $m_{BH,tot}$ for all models.
Top and bottom panel show the results of triple- and two-BH models respectively.
 \label{fig6}}
\end{figure}

Figure \ref{fig6} shows the ratio of mass deficit to total mass of BHs
for all models.
Here, we define the mass deficit as difference of mass at any given time 
with original mass within ``break radius'',
defined as the radius at which the initial profile 
and that at the current time intersect with each other
(see Figure \ref{fig2}).
In the case of binary BH runs (model MxB),
$m_{{\rm def}}/m_{{\rm BH,tot}}$ reaches about unity by $T \sim 1$,
and the increase after that is due to loss-cone refilling 
by relaxation.
Therefore, in the case where relaxation is negligible,
$m_{{\rm def}} \sim m_{{\rm BH,tot}}$.
In the case of triple BH system, however, $m_{{\rm def}}/m_{{\rm BH,tot}}$ does tell
about the past history of the evolution of the triple system.
Thus, a galaxy with unusually large $m_{{\rm def}}/m_{{\rm BH,tot}}$ is a good candidate for
the host of triple BH.

In Figure 7, 
we compare the two methods to estimate the mass deficit,
one of which is our definition and the other is used by
MM2001.
The estimate mass from MM2001 is 
systematically higher than real one.
This difference is simply due to the fact that 
the method of MM2001 overestimates the initial mass 
with the break radius.
We tried to use the method proposed by \citet{Graham04}, 
but it turned out that the fit by a S\'{e}rsic profile
is not appropriate for our galaxy model.

\begin{figure}
\plotone{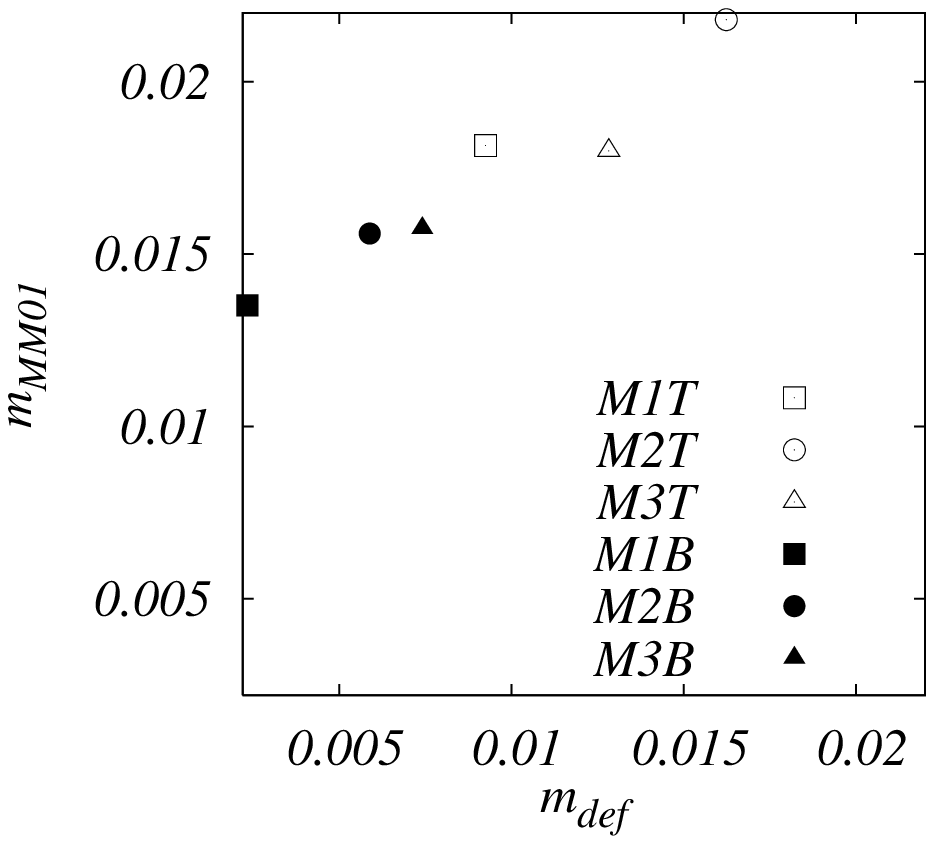}
 \caption
 {
Mass deficit estimated with MM2001's methods and 
the real value.
For MxB runs, $m_{{\rm def}}$ is calculated at $T=1$,
and for MxT runs at $T=3$.
 \label{fig7}}
\end{figure}

\subsection{Velocity Structure}

\begin{figure}
\plotone{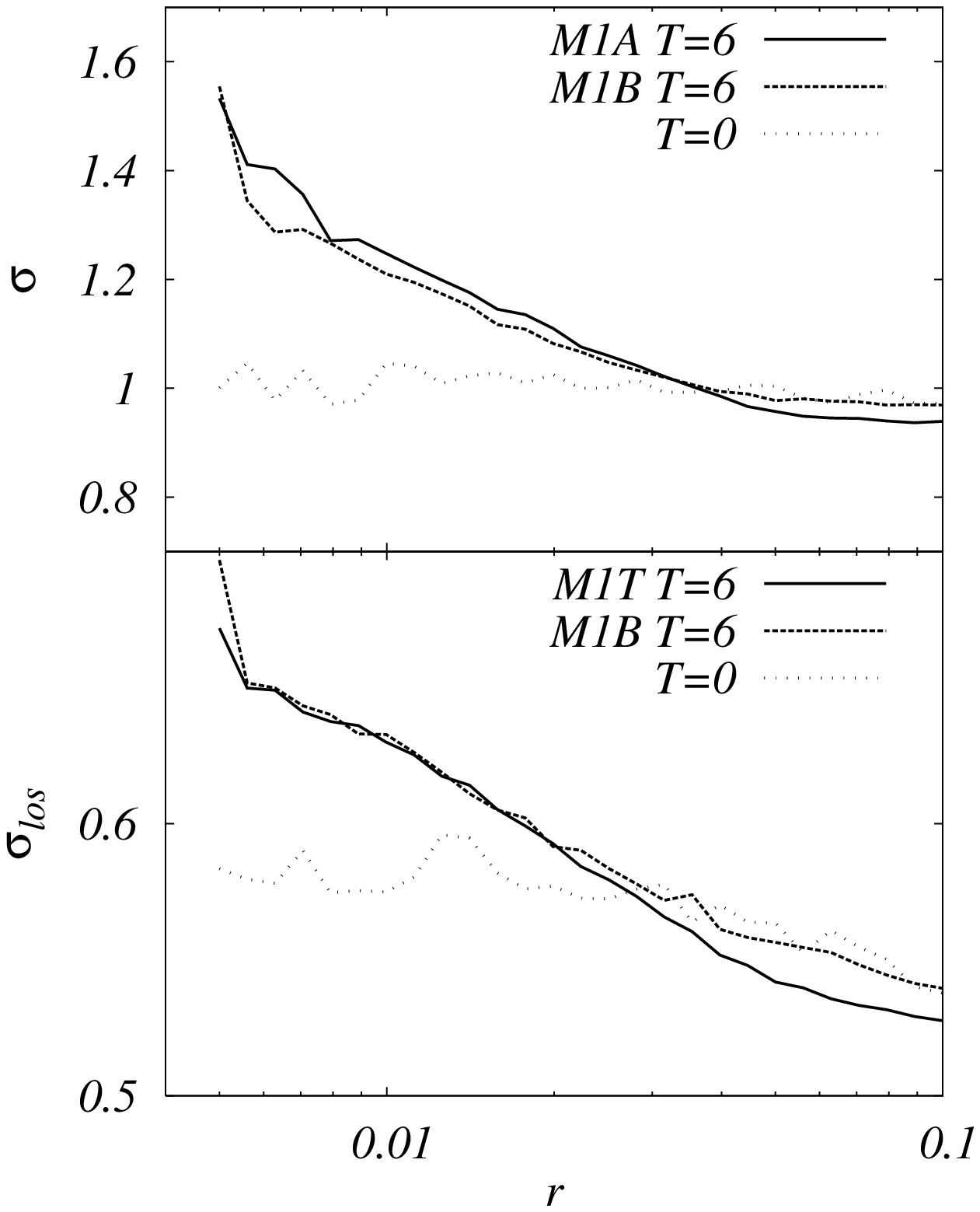}
 \caption
 {
Velocity dispersion (top) and line-of-sight velocity (bottom)
profiles at $T=6$ for models M1T and M1B.
Solid and dashed curves show three- and two-BH cases respectively,
and dotted curves show the initial profiles.
 \label{fig8}}
\end{figure}

Figure \ref{fig8} shows the velocity dispersion (top) and
line-of-sight velocity (bottom) profiles at $T=6$.  In the
central region, the velocity dispersion is increasing
toward the center.  Since the potential of BHs dominates the potential
in the central region, the orbits of stars are almost Keplerian.
The line-of-sight velocity dispersion shows the similar rise at the
center. 

\begin{figure}
\epsscale{0.8}
\plotone{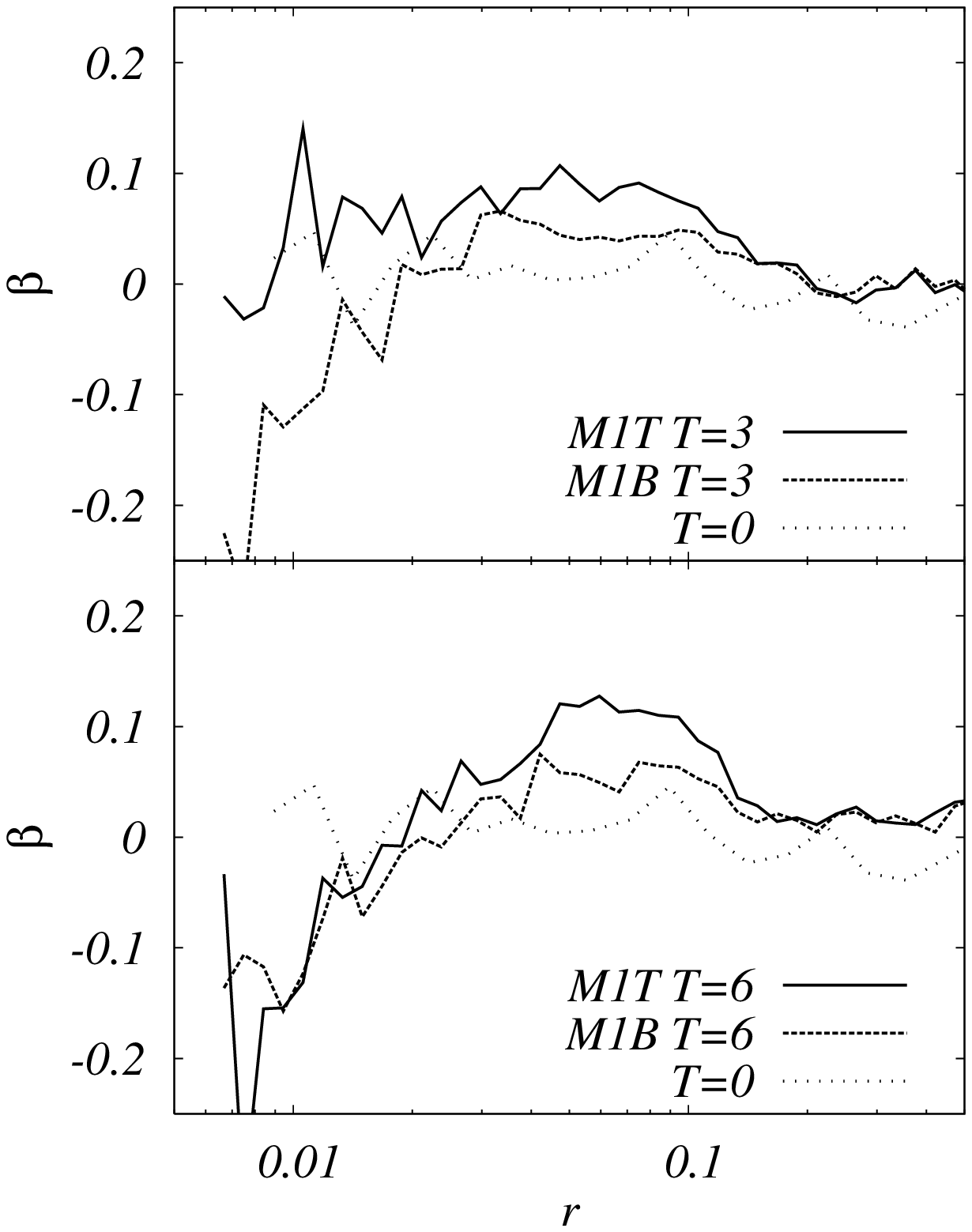}
 \caption
 {Anisotropy parameter 
 for models M1T and M1B.
 Solid and dashed curves show the results of 
 three- and two-BH cases respectively, and
 dotted curves show the initial profiles.
 \label{fig9}}
\end{figure}

In Figure \ref{fig9}, the velocity anisotropy for models M1T and
M1B are compared.  Here the anisotropy parameter is defined as
\begin{equation}
\beta \equiv 1-\frac{\langle v_t^2 \rangle}{2\langle v_r^2 \rangle},
\end{equation}
where $\langle v_t^2 \rangle$ and $\langle v_r^2 \rangle$ are the
tangential and radial mean square velocities.  For model M1B, orbits
are circular in the central region and $\beta$ increases towards 0
as $r$ increases at $T=3$ and $6$.  From figures
\ref{fig3} and \ref{fig9}, it is clear that, after two BHs
settled and formed a binary in the center, the orbits of stars in the
cusp becomes tangentially anisotropic  and in the outer region 
the orbits remain  
isotropic.  This is because the BH binary kicked out stars with radial
orbits more efficiently than those with circular orbits in inner region.

For run M1T, $\beta$ around the center is close to zero (isotropic)
at $T=3$, while that in outer region is positive.  Thus,
that stars are  radially anisotropic around the outer edge of the
central cusp,  in the triple BH 
case.
In this model, $\beta$ around the center decreases after
merging at $T=6$ like the two BH case. This is because of the slow
evolution of the binary BH through the interaction with field stars.

\section{DISCUSSION AND CONCLUSION}

\subsection{Possibility to Find Triple BH Systems}

In previous sections, we have seen that the density and velocity
structure of galaxies with triple BHs is different from that with
two BHs.  In Paper I, we showed that two of three BHs in a galaxy merge
within several dynamical times.  The standard hierarchical clustering
scenario of galaxy formation suggests that galaxies frequently merge
and form ellipticals.  Many elliptical galaxies, therefore, might host a
binary BH. It is likely that many of them had a triple BH system once
upon a time.
Furthermore, there may be galaxies with a triple BH system.
The lifetime of a triple BH system in a galaxy is 
several times the dynamical
time of the core of the galaxy, which is  about $10^8$ 
years ($100$ M years).
If we assume every elliptical galaxy have binary BH,
and all large ellipticals are formed through mergings
of elliptical galaxy.
Speaking, 1\% of
large ellipticals could have triple BH systems now.

One possible way to discriminate between a galaxy with binary or
single BH and that have (or had) triple BH is the measurement of the
cusp radius and density.  Our results suggest that the central cusp of
a galaxy with three BHs is larger by a factor of few than that with
binary BH.  Thus the galaxy with a large cusp and low density in the
central region might have (or had) three BHs.

Another possible way is to use the difference of velocity anisotropy.
Anisotropy parameter of galaxy with three BHs
is positive (radial) in the outer region 
due to the heating by BHs.

The anisotropy parameter $\beta$ is estimated
by the line of sight velocity and their higher moment,
assuming that the galaxy is spherical.
However, with this method it is difficult to
obtain local anisotropy parameter in the central region of galaxies,
unless the central density cusp is steep \citep{Gerhard93}.
Another method is to fit the model with observational data
\citep{Cretton00,Gebhardt00,Gebhardt03}.
Figure 10 in \citet{Gebhardt03}
shows anisotropy profile of galaxies.
The shapes of some galaxies profile
in their paper,
for example all ``weak-cusp'' galaxies (NGC3608, NGC4291, NGC4649),
are similar to of our result in Figure \ref{fig9}.
These galaxies might contain or have contained triple BHs.

\subsection{Conclusions}

In this paper, we investigated the effect
of two- or three-BH systems on the structure of galaxies.
We found that  if the galaxy contains three BHs, 
(1) multiple three body scattering events reduce
the density of the central cusp 
and increase the cusp radius
and (2) in outer region, orbits of stars are likely to be radial.
These difference allow us to
discriminate the galaxies with two or three BHs.

\acknowledgments
We thank Toshiyuki Fukushige,
Yusuke Tsukamoto and Keigo Nitadori 
for stimulating discussions and useful comments.

This research is partially supported by the Special Coordination Fund
for Promoting Science and Technology (GRAPE-DR project), Ministry of
Education, Culture, Sports, Science and Technology, Japan.

\appendix


\begin{thebibliography}{}

\bibitem[Begelman, Blandford, \& Rees(1980)]{bbr80} 
  Begelman, M.~C.,  Blandford, R.~D., \& Rees, M.~J. 1980, \nat, 287, 307


\bibitem[ Berczik, Merritt, \& Spurzem(2005)]{berczik2005}
Berczik, P., Merritt, D., \& Spurzem, R. 2005, \apj, 633, 680

\bibitem[Binney \& Tremaine(1987)]{GD}
  Binney, J., \& Tremaine, S. 1987, Galactic Dynamics (Princeton:
					   Princeton Univ. Press)

\bibitem[Blas, Lee, \& Socrates (2002)]{BLS}
Blas, O., Lee, M.H., \& Socrates, A. 2002 \apj, 578, 775

\bibitem[Casertano \& Hut(1985)]{CH85}
{Casertano}, S., \& {Hut}, P. 1985, ApJ, 298, 80

\bibitem[Cretton et al.(2000)]{Cretton00}
 Cretton, N., Rix, H.-W., \& de Zeeuw, P.~T.\ 2000, \apj, 536, 319 

 \bibitem[Damour(1987)]{Damour87}
Damour, T. 1987, in Three Hundred Years of Gravitation, ed. S. Howking
					   \& W.Israel (Cambridge:
					   Cambridge Univ. Press), 128

\bibitem[{{Ebisuzaki} {et~al.}(1991){Ebisuzaki}, {Makino}, \&
  {Okumura}}]{Ebisuzakietal1991}
{Ebisuzaki}, T., {Makino}, J., \& {Okumura}, S.~K. 1991, Nature, 354, 212

\bibitem[Faber et al.(1997)]{faber97}
Faber, S.M., et al. 1997, \aj, 114, 1771

\bibitem[Gebhardt et al.(2000)]{Gebhardt00}
 Gebhardt, K., et al.\ 2000, \aj, 119, 1157

\bibitem[Gebhardt et al.(2003)]{Gebhardt03}
Gebhardt, K., et al.\ 2003, \apj, 583, 92 

\bibitem[Gerhard(1993)]{Gerhard93}
Gerhard, O.~E.\ 1993, \mnras, 
265, 213 

\bibitem[Graham(2004)]{Graham04}
Graham, A.~W.\ 2004, \apjl, 613, L33 

\bibitem[Heggie \& Mathieu(1986)]{Hg}
Heggie, D.~C., \& Mathieu, R.~D. 1986, in The Use of Supercomputers in
					   Stellar Dynamics, ed. P. Hut
					   \& S. McMillan (Berlin:
					   Springer), 233
\bibitem[Hoffman \& Loeb (2007)]{HL07}
Hoffman, L., \& Loeb, A. 2007, \mnras, 377, 957

\bibitem[Iwasawa, Funato \& Makino (2006)]{IFM06}
Iwasawa, M., Funato, Y., \& Makino, J. 2006, \apj, 651, 1059

\bibitem[King(1966)]{kin66}  King, I. R.  1966, \aj, 71, 276

\bibitem[Kormendy \& Richstone(1995)]{KR95}
 Kormendy, J., \& Richstone, D.\ 1995, \araa, 33, 581 

\bibitem[Kozai(1962)]{Kozai62} 
 Kozai, Y. 1962, \aj, 67, 591

\bibitem[Lauer et al.(1995)]{lauer95} 
Lauer, T. R., et al.  1995, \aj, 110, 2622

\bibitem[Magorrian et al.(1998)]{magorrian98} 
Magorrian, J., et al. 1998, \aj, 115, 2285

\bibitem[Makino \& Aarseth(1992)]{her} 
Makino, J. \&  Aarseth, S. 1992, \pasj, 44, 141

\bibitem[Makino \& Ebisuzaki(1994)]{ME94} 
Makino, J. \&  Ebisuzaki, T. 1994, \apj, 436, 607

\bibitem[{{Makino} \& {Ebisuzaki}(1996)}]{ME96}
---. 1996, \apj, 465, 527

\bibitem[Makino \& Funato(2004) ]{MF04} 
Makino, J. \& Funato, Y. 2004, \apj, 602, 93

\bibitem[Makino et al.(2003)]{grape6} 
Makino, J., Fukushige, T., Koga, M.  \&  Namura, K., 2003, \pasj, 55, 1163

\bibitem[Marconi \& Hunt(2003)]{marconi03} 
Marconi, A., \& Hunt, L. K. 2003, \apj, 589, 21

\bibitem[Merritt(2006)]{merritt06} 
Merritt, D. 2006, \apj, 648, 976

\bibitem[{{Milosavljevi{\' c}} \& {Merritt}(2001)}]{MM01}
{Milosavljevi{\' c}}, M., \& {Merritt}, D. 2001, ApJ, 563, 34

\bibitem[Nakano \& Makino (1999)]{NM99}
Nakano, T. \& Makino, J. 1999, \apj, 510, 155


\end{thebibliography}
\end{document}